\documentclass[acmtog,nonacm,screen]{acmart}

\usepackage{enumitem}
\usepackage{xcolor}
\usepackage[normalem]{ulem}
\newcommand{\xx}{\mathbf{x}}

\newcommand{\vvv}{\mathbf{v}}
\newcommand{\ii}{\mathbf{i}}

\newcommand{\dd}{\mathbf{d}}
\newcommand{\pp}{\mathbf{p}}

\renewcommand{\ss}{\mathbf{s}}
\renewcommand{\ll}{\mathbf{l}}

\newcommand{\MM}{\mathbf{M}}

\newcommand{\PP}{\mathbf{P}}

\renewcommand{\ll}{\mathbf{l}}

\newcommand{\JJ}{\mathbf{J}}

\newcommand{\nn}{\mathbf{n}}

\newcommand\restr[2]{{\left.\kern-\nulldelimiterspace{}#1\right|_{#2}}}

\newcommand{\oo}{\mathbf{o}}

\usepackage[ruled]{algorithm2e} 

\SetAlFnt{\small}
\SetAlCapFnt{\small}
\SetAlCapNameFnt{\small}
\SetAlCapHSkip{0pt}

\newcommand{\TODO}[1]{{\color{red}{TODO: #1}}}

\acmJournal{TOG}

\begin{document}
\title{Strand-based Hairstyle Generation via  Large Reconstruction and Multimodal Models}

\author{Conghui Hao}
\orcid{0009-0004-1956-4364}
\email{free7187freia@gmail.com}
\affiliation{%
  \institution{LIGHTSPEED}
  \city{Shenzhen}
  \country{China}
}

\author{Tao Huang}
\orcid{0009-0002-3458-0851}
\email{tao_huang@ucsb.edu}
\affiliation{%
  \institution{LIGHTSPEED}
  \city{Shenzhen}
  \country{China}
}

\author{Yuefan Shen}
\orcid{0000-0002-6049-7966}
\email{yuefanshen@outlook.com}
\affiliation{%
  \institution{LIGHTSPEED}
  \city{Shenzhen}
  \country{China}
}

\author{Tongtong Wang}
\orcid{0009-0005-6585-3009}
\email{wangtong923@gmail.com}
\affiliation{%
  \institution{LIGHTSPEED}
  \country{Australia}
}

\author{Zhongtian Zheng}
\orcid{0009-0009-4714-1760}
\email{zhengzhongtian@pku.edu.cn}
\affiliation{%
  \institution{LIGHTSPEED}
  \city{Shenzhen}
  \country{China}
}

\author{Kui Wu}
\orcid{0000-0003-3326-7943}
\email{kwwu@lightspeed-studios.com}
\affiliation{%
  \institution{LIGHTSPEED}
  \city{Los Angeles}
  \state{CA}
  \country{USA}
}



\begin{abstract}
Creating high-quality strand-based hairstyles in current production pipelines remains heavily dependent on skilled artists and time-consuming manual authoring, making it costly and difficult to scale. Existing learning-based methods have advanced image-driven hair reconstruction, but typically require large, diverse training datasets, struggle to generalize to complex styles such as buns and ponytails, and often operate in representations that are not directly compatible with strand-based modeling, editing, and simulation. 
We present a novel automatic pipeline that combines the capabilities of Large Reconstruction Models (LRMs), Large Multimodal Models (LMMs), and classical geometry processing to generate high-quality strand-based hairstyles from single-view images. Our approach produces detailed, production-ready strand geometry without task-specific training or data collection and can handle a wide variety of hairstyles, including straight and curly hair, short and long styles, and challenging structured configurations such as ponytails and buns. Across this diverse set of examples, our method generates visually compelling strand-level reconstructions within only a few minutes, making it well-suited for integration into modern digital human workflows.
\end{abstract}

%
%


%
%
\begin{teaserfigure}
\centering
\includegraphics[width=\textwidth]{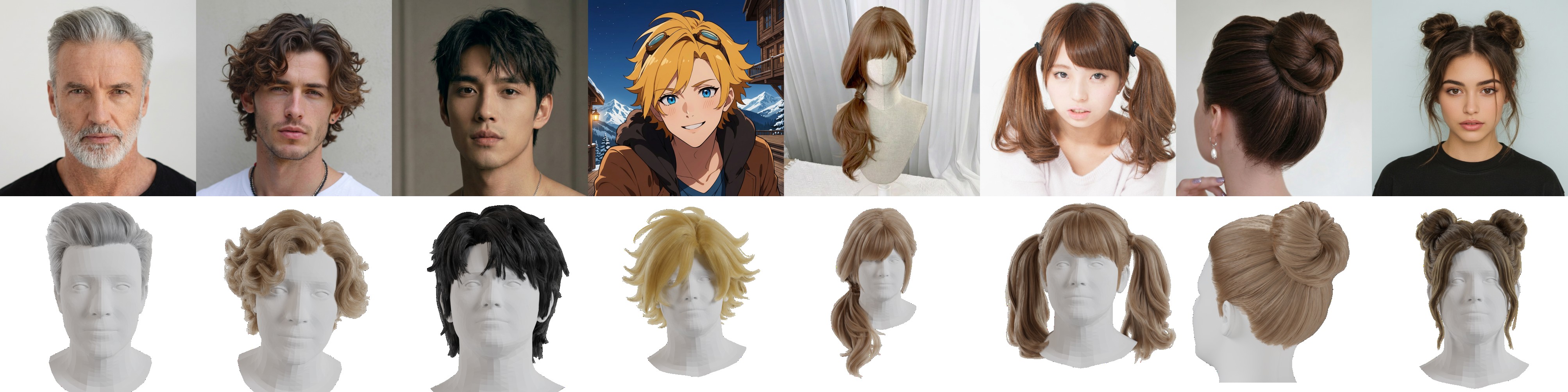}
\captionof{figure}{Given an input image, our method combines the capabilities of Large Reconstruction Models (LRMs), Large Multimodal Models (LMMs), and classical geometry processing to automatic generate high-quality strand-based hairstyles with a wide range of hairstyle complexities, including straight, wavy, and curly hair, short and long styles, as well as challenging cases such as buns and ponytails.
}
\Description{}
\label{fig:teaser}
\end{teaserfigure}


\maketitle
\section{Introduction}

Hair modeling and generation are essential to a broad range of applications involving digital humans, including avatars, augmented and virtual reality, games, film production, and telepresence. As one of the most visually salient and identity-defining components of human appearance, hair plays a central role in perceived realism, personalization, and stylistic diversity. Despite its importance, creating high-quality hairstyles in existing industrial pipelines still relies heavily on skilled artists and labor-intensive manual authoring, making the process costly and difficult to scale. 

Recent learning-based methods have made significant progress in 3D hair reconstruction and generation from images. Many existing methods represent hair using neural networks that predict volumetric orientation fields~\cite{zheng2023hairstep,wu2022neuralhdhair,wu2024monohair}, from which individual strands are later traced as a post-processing step. While effective, these approaches depend heavily on large, diverse training datasets to capture generalizable priors. This limitation becomes particularly pronounced for complex hairstyles, such as ponytails and buns, which are relatively rare in available datasets. Another line of work uses variational autoencoders (VAEs) to encode strand geometry and global hairstyle structure into compact latent representations~\cite{difflocks2025,zhou2023groomgen,he2025perm,zhou2024groomcap,sklyarova2024text,sklyarova2025im2haircut}. Although such models are well suited for capturing overall style and coarse structure, their compact embeddings often lack the expressive power needed to recover the full geometric complexity of real hair. More broadly, like other learning-based approaches, they also struggle to generalize to complex hairstyles that are underrepresented in training data.  Several methods further improve visual plausibility through differentiable rendering~\cite{takimoto2024dr,rosu2022neural,sklyarova2023neural,zhou2024groomcap}. 
However, image-space supervision provides limited geometric constraints due to the inherent 2D-to-3D ambiguity caused by severe self-occlusions of strands. 
Consequently, despite their promise, existing learning-based methods remain limited in their ability to reconstruct high-quality, highly detailed hairstyles.
Recent advances in deep learning have shown that Large Multimodal Models (LMMs)~\cite{nanobanana2026,openai2024gpt4o,midjourney2026} are highly effective at understanding and describing images, while Large Reconstruction Models (LRMs)~\cite{kerbl20233d, mildenhall2021nerf, lai2025hunyuan3d} can generate high-fidelity 3D surfaces. However, neither image-space information nor surface-manifold representations can be directly used as strand-based geometry for realistic hair modeling, editing, rendering, and simulation in current 3D production pipelines.

In this work, we present a novel, fully automatic pipeline that combines the capabilities of LMMs and LRMs with conventional graphics techniques to generate high-quality strand-based hairstyles. In particular, our method first reconstructs a hair surface mesh from the input image using an LRM, together with a bust template and scalp mask. We then render multiple surrounding views and leverage an LMM to generate line art that serves as structural guidance for estimating hair directions in image space, which are then projected onto the hair surface. Next, LMMs are used to identify parting lines and segment the hairstyle into gathering, constrained, loose, and scalp regions. A three-stage diffusion-correction-diffusion process is performed to complete the surface directions, resolve orientation ambiguity, and propagate the oriented field into the volume. Finally, we generate hair strands by tracing through the recovered orientation field.
We evaluate our method across a wide range of hairstyle complexities, including straight and curly hair, short and long styles, and challenging cases such as buns and ponytails. Compared with previous methods, our approach requires neither training nor data collection, while supporting a broader range of hairstyles and generating high-quality results efficiently within only a few minutes.

\section{Related Work}

\paragraph{Classical Multi-view Hair Acquisition.}
Early work on hairstyle acquisition from multi-view imagery~\cite{Grabli2002,Paris2004,Wei2005} largely followed an analysis-by-synthesis paradigm. These methods first estimated local 2D hair orientations from images using oriented filter responses, and then synthesized 3D strands by growing fibers inside a reconstructed hair volume. Since orientation cues were available only in visible regions, subsequent work focused on inferring the unobserved structure. \citet{paris2008hair} propagated structure tensors throughout the volume to infer missing fiber directions and obtain a more complete orientation field. \citet{jakob2009capturing} further improved fine-scale capture by using many photographs with sweeping focal planes and applying Gabor filters to better detect strand features across depth. Our work follows a similar idea of structure tensor propagation, but leverages LMMs for more reliable semantic labeling and stable results.
Later methods sought to reconstruct plausible 3D hair geometry under more limited input conditions. \citet{Luo2012} introduced structure-aware aggregation to enforce continuity implied by local hair orientations, followed by template refinement to recover globally consistent geometry; this line of work was later extended to sparse wide-baseline camera arrays~\cite{Luo2013}. \citet{Luo2013hair} also proposed clustering local strand segments and bridging gaps in point clouds, while \citet{Hu2014} incorporated physics-based simulation to improve structural plausibility. \citet{Nam2019} instead considered reconstructing accurate strand segments from noisy point-based observations and proposed a mean-shift-based line denoising procedure to convert unorganized measurements into strand segments.

\paragraph{User-guided and Database-driven Modeling.}
To resolve directional ambiguity that is difficult to infer from images alone, several methods incorporate user supervision. \citet{Chai2012} showed that approximate strand models can be recovered from a single image by exploiting inter-strand occlusion relationships via labeling. \citet{Chai2013} used sparse user annotations to guide the recovery of a dense, globally consistent 3D hair direction field via linear least-squares optimization. \citet{Zhang2017} adopted a similar strategy to construct volumetric direction fields for hair reconstruction. 
Another important direction improves reconstruction quality by leveraging hairstyle priors from databases. \citet{Wang09} synthesized new hairstyles from guide strands using ideas from texture synthesis. \citet{Hu2014braid} reconstructed braided hairstyles by matching observations to a family of procedural braid models, while \citet{hu2015single} retrieved exemplar hairstyles that best matched user-provided 2D strokes. \citet{chai2016autohair} classified an input into a small set of spatial hair distribution classes, then predicted segmentation and growth direction maps, and finally retrieved the most compatible 3D exemplar from a large database. \citet{Zhang2018} further introduced patch-level local similarity for exemplar search and used retrieved examples to synthesize a complete 3D orientation field. Recently, \citet{Meishvili2024} revisited this paradigm with a classification-based predictor that selects the closest hairstyle from a dataset given a single image.

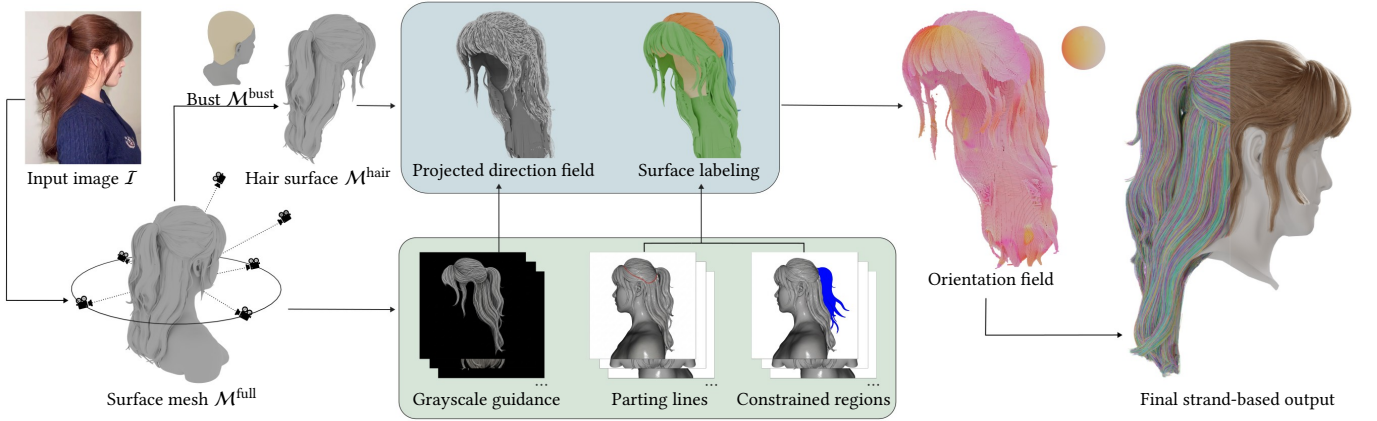
\begin{figure*}[t!]
\centering
\includegraphics[width=\linewidth]{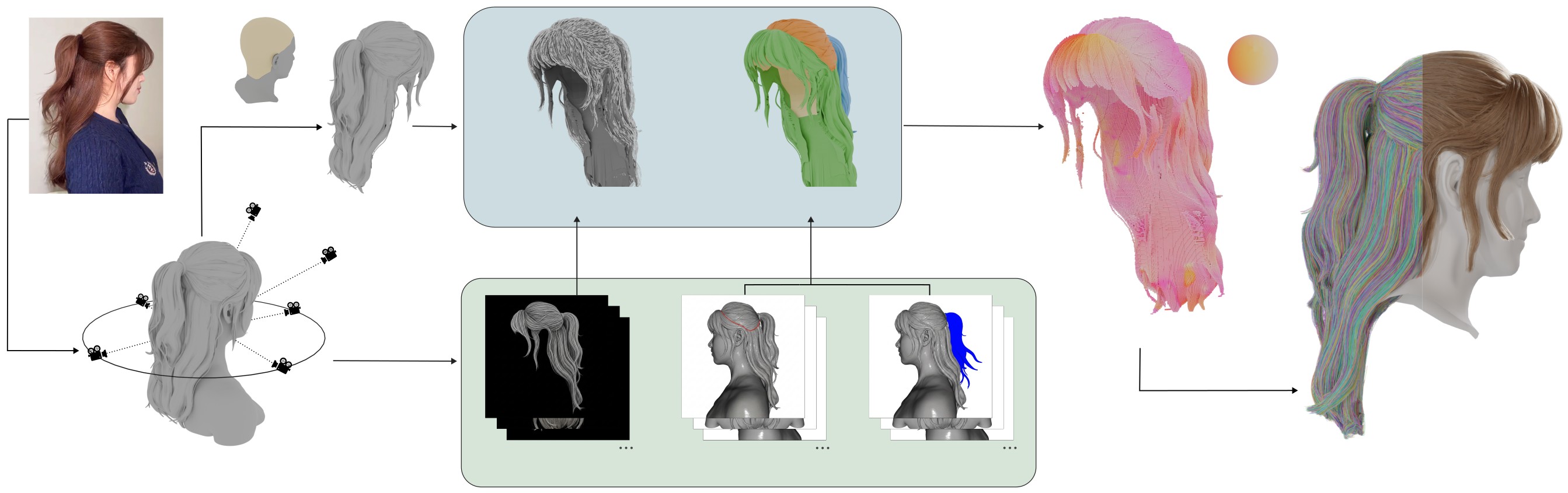}
\put(-500,90){\footnotesize Input image $\mathcal{I}$}
\put(-470,7){\footnotesize Surface mesh $\mathcal{M}^{\text{full}}$}
\put(-418,90){\footnotesize Hair surface $\mathcal{M}^{\text{hair}}$}
\put(-440,120){\footnotesize Bust $\mathcal{M}^{\text{bust}}$}
\put(-355,93){\footnotesize Projected direction field}
\put(-270,93){\footnotesize Surface labeling}
\put(-355,7){\footnotesize Grayscale guidance}
\put(-280,7){\footnotesize Parting lines}
\put(-233,7){\footnotesize Constrained regions}
\put(-161,52){\footnotesize Orientation field}
\put(-81,7){\footnotesize Final strand-based output}
\caption{The pipeline of our framework. From a single input image $I$, we reconstruct a hair mesh $\mathcal{M}^{\text{hair}}$ via an LRM and query an LMM across multiple views to obtain semantic cues, including direction guidance, parting lines, and constraint regions, which segment the hair surface into multiple regions based on their structure. All visual cues are projected onto the surface and used to recover a globally consistent orientation field by solving a three-stage orientation diffusion process, which is then applied to strand tracing to generate the final strand-based output.} 
\label{fig:pipeline}
\Description{}
\end{figure*}

\paragraph{Neural Representations and Differentiable Rendering.}
Recent work increasingly relies on neural hair representations together with differentiable rendering for reconstruction and refinement. Neural Haircut~\cite{rosu2022neural} represented hair using a neural scalp texture and optimized it with a neural rendering. \citet{sklyarova2023neural} reconstructed implicit surfaces for the bust and hair, learned a 3D orientation field, and improved fidelity with a differentiable hair renderer. \citet{Zakharov2024} introduced a dual representation based on strand polylines and 3D Gaussians to enable differentiable rasterization. More recently, DiffLocks~\cite{difflocks2025} leveraged a large-scale synthetic strand dataset to train an image-conditioned diffusion transformer for single-image 3D strand generation, while Im2Haircut~\cite{sklyarova2025im2haircut} combined full 3D supervision on synthetic data with self-supervised losses on real images.

\paragraph{Learning-based Strand and Prediction.}
Deep learning has substantially advanced hair reconstruction and generation from sparse visual input. HairNet~\cite{zhou2018hairnet} pioneered a strand-based encoder--decoder architecture that maps a 2D orientation field to strands distributed over a parameterized scalp. Subsequent work extended learning-based hair modeling to dynamic capture from monocular video~\cite{Yang2019}, sketch-based modeling~\cite{shen2020deepsketchhair}, and CT-based reconstruction~\cite{Shen2023ct2hair}. Several methods explicitly model hairstyles through latent spaces. \citet{zhou2023groomgen} introduced a strand VAE for individual strand geometry together with a hairstyle VAE for sparse guide hairs, while \citet{he2025perm} used a PCA-based frequency-domain strand representation to improve disentanglement and editing control. GroomCap~\cite{zhou2024groomcap} instead learned a neural implicit volumetric representation of occupancy and orientation and further refined the result with Gaussian-based strand optimization. Recently, \citet{luo2026} reformulated hair strand generation as a dual decoupled autoregressive process and introduced a geometric tokenization scheme for generating guide strands. Compared with recent learning-based methods, our approach produces detailed, production-ready strand geometry without any task-specific training or data collection and can handle a wide variety of hairstyles more efficiently.

\paragraph{Volumetric and Orientation-field Representations.}
A parallel line of work represents hair as volumetric occupancy and orientation fields. \citet{Saito2018} learned hairstyles with a volumetric VAE from a compact latent code. NeuralHDHair~\cite{wu2022neuralhdhair} proposed a coarse-to-fine network for predicting a high-fidelity 3D orientation field from a single image, and DeepMVSHair~\cite{kuang2022deepmvshair} extended this formulation to sparse-view input by inferring occupancy and growth direction from multi-view pixel-aligned features rather than optimizing a shape prior as in~\cite{Zhang2017}. MonoHair~\cite{wu2024monohair} combined a coarse geometry initialization from neural radiance fields with video-based aggregation to recover detailed hair exteriors, and then estimated volumetric occupancy and orientation fields. Simultaneously, \citet{takimoto2024dr} combined 3D orientation estimation with global optimization based on Laplace's equation. HairStep~\cite{zheng2023hairstep} similarly predicted directed 2D orientation maps and depth from a single view, then lifted them to 3D occupancy and orientation representations.  Recently, \citet{shen2026HairLRM} employed a Dual Orientation AutoEncoder to lift coarse geometry into high-fidelity hair strands conditioned on a hair mesh generated by a large reconstruction model (LRM). However, purely learning-based approaches, such as HairLRM, fundamentally rely on large and diverse training datasets and may struggle with out-of-distribution hairstyles. Similarly, our method targets the reconstruction of strand-based hair models from single- or multi-view image inputs and utilizes meshes reconstructed by Large Reconstruction Models as intermediate geometric representations. However, instead of learning strand generation, we leverage Large Multimodal Models to predict hair segmentation and orientation on the reconstructed mesh, followed by a three-stage procedure for propagating and optimizing directions using classical geometry-processing techniques.

\section{Method}

We present a novel framework for reconstructing high-quality strand-based hair models with complex structures from a single image by combining large reconstruction models (LRMs) and large multimodal models (LMMs). As illustrated in \autoref{fig:pipeline}, our pipeline begins by reconstructing a coarse 3D surface from the input portrait image using an LRM. After aligning the reconstruction with a template bust model, we extract a watertight hair mesh and its associated scalp region (\autoref{sec:image2mesh}). We then render the reconstructed surface from multiple surrounding viewpoints and use an LMM to infer image-space hair directions, which are subsequently projected onto the hair surface (\autoref{sec:projection}).
%
An LMM is used to identify parting lines and constrained structures, and partition the hairstyle into loose, gathering, constrained, and scalp regions (\autoref{sec:partition}). We next sample the interior of the hair volume and construct a graph over both volumetric samples and surface vertices. On this graph, we perform a three-stage diffusion-correction-diffusion procedure to propagate the oriented field into the hair volume (\autoref{sec:diffusion}). The recovered volumetric orientation field is then used to trace hair strands (\autoref{sec:tracing}).
Throughout the paper, we use \emph{direction} $\dd$ to denote an \emph{unoriented} unit vector and \emph{orientation} $\oo$ to denote an \emph{oriented} unit vector.

\subsection{Image-to-Mesh}\label{sec:image2mesh}

Given an input portrait image $\mathcal{I}$, we first apply a state-of-the-art image-to-3D large reconstruction model (LRM) to obtain an initial surface mesh $\mathcal{M}^{\text{full}}$. We identify the hair region based on texture cues and retain only its largest connected component. Then, we align the reconstructed mesh to a template bust model~\cite{li2017learning}, denoted by $\mathcal{M}^{\text{bust}}$, which is equipped with a scalp mask. The alignment is performed in two stages. We first compute a coarse alignment by matching the bounding boxes of $\mathcal{M}^{\text{full}}$ and $\mathcal{M}^{\text{bust}}$, thereby estimating a global scale and translation. We then refine the alignment using 2D facial landmarks detected from rendered views of the reconstruction. These landmarks are back-projected into 3D and further adjusted using the inter-pupillary distance (IPD) as a stable metric for facial scale.
After alignment, we subtract the bust mesh from the full reconstruction to obtain a watertight hair mesh $
\mathcal{M}^{\text{hair}} = \mathcal{M}^{\text{full}} \backslash \mathcal{M}^{\text{bust}}.$ We denote the enclosed hair volume by $\Omega$. The scalp mask inherited from the aligned bust model will later provide both semantic guidance and boundary conditions for orientation diffusion.

\subsection{Direction Projection} \label{sec:projection}

Estimating hair direction from a single image is challenging. Prior approaches rely on manual annotation~\cite{Chai2013,hu2015single} 
or learned predictors~\cite{zhou2018hairnet,zheng2023hairstep}. In practice, however, these approaches are often unreliable in our setting: manual labeling is tedious and difficult to scale 
and learned predictors are often limited by training bias and insufficient viewpoint coverage. Our key observation is that LMMs can provide globally coherent visual guidance (\autoref{fig:pipeline}), particularly useful for inferring plausible hair flow from sparse image evidence. 


To exploit this capability, we render the aligned full mesh $\mathcal{M}^{\text{full}}$ from six viewpoints: front, left, back, right, and two additional top-down frontal views from the left and right. These additional views provide stronger cues for global flow and overall hairstyle structure. This produces a set of rendered images $\{ \mathcal{I}^{\text{render}}_i \}$ at resolution $1024 \times 1024$. To avoid misleading cues from the unstable LRM textures, we render the mesh without texture. Instead, for each rendered image, an LMM generates a grayscale guidance image $\mathcal{I}^{\text{gray}}_i$ whose intensity variations emphasize the dominant local strand flow.
Given a guidance image $\mathcal{I}^{\text{gray}}_i$, we estimate the local image-space direction at each pixel $\ii$ inside the hair region using the 2D structure tensor. Specifically, we compute image gradients $\partial \mathcal{I}_x(\ii)$ and $\partial \mathcal{I}_y(\ii)$ and form
\begin{align}
\JJ =
\begin{pmatrix}
\langle \partial \mathcal{I}_x^2 \rangle & \langle \partial \mathcal{I}_x \, \partial \mathcal{I}_y \rangle \\
\langle \partial \mathcal{I}_x \, \partial \mathcal{I}_y \rangle & \langle \partial \mathcal{I}_y^2 \rangle
\end{pmatrix},
\end{align}
where $\langle \cdot \rangle$ denotes Gaussian smoothing with kernel size $19$. The eigenvectors of $\JJ$ define the principal directions of local variation, and the eigenvalues quantify anisotropy. Since hair strands appear as locally elongated line structures, we take the eigenvector corresponding to the smaller eigenvalue as the local image-space hair direction, denoted by $\dd^{\text{image}}$.

Based on the camera intrinsics and extrinsics for each rendered view, $\dd^{\text{image}}$ is lifted to a 3D world-space direction $\dd^{\text{world}}$ . For each pixel, we cast the corresponding camera ray $\ll$ to find its intersection with the hair mesh $\mathcal{M}^{\text{hair}}$. Let $\nn$ denote the normal of the intersected face. We project $\dd^{\text{world}}$ onto the tangent plane of the face via
\begin{align}
\dd^{\text{proj}}
=
\dd^{\text{world}}
-
{(\dd^{\text{world}} \cdot \nn)} / {(\ll \cdot \nn)}\,\ll.
\end{align}
Because the recovered field is still unoriented, directly averaging vectors is invalid: $\dd$ and $-\dd$ represent the same direction and may cancel each other numerically. Following~\cite{Paris2004}, we represent each direction using the sign-invariant symmetric tensor
\begin{align}
\MM = \dd \dd^{\top}.
\end{align}
For each surface vertex, we aggregate all projected tensors from nearby projected samples on adjacent faces across all views using inverse-distance weighting:
\begin{align}
\MM^{\text{vertex}}
=
{
\sum_i \frac{1}{\|\xx_i^{\text{proj}} - \xx^{\text{vertex}}\|} \, \MM_i^{\text{proj}}
}/ {
\sum_i \frac{1}{\|\xx_i^{\text{proj}} - \xx^{\text{vertex}}\|}
},
\end{align}
where $\xx_i^{\text{proj}}$ is the position of projected sample $i$ and $\xx^{\text{vertex}}$ is the vertex position. We recover the vertex direction $\dd^{\text{vertex}}$ by eigendecomposing $\MM^{\text{vertex}}$ and taking the principal eigenvector.

\subsection{Semantic Partition}\label{sec:partition}

The projected surface directions provide local geometric cues, but they are insufficient to resolve the global organization of many hairstyles. 
To reduce this ambiguity, we extract a small set of semantic cues that strongly constrain the underlying hair flow.

\begin{figure}[ht!]
    \centering
    \includegraphics[width=\linewidth]{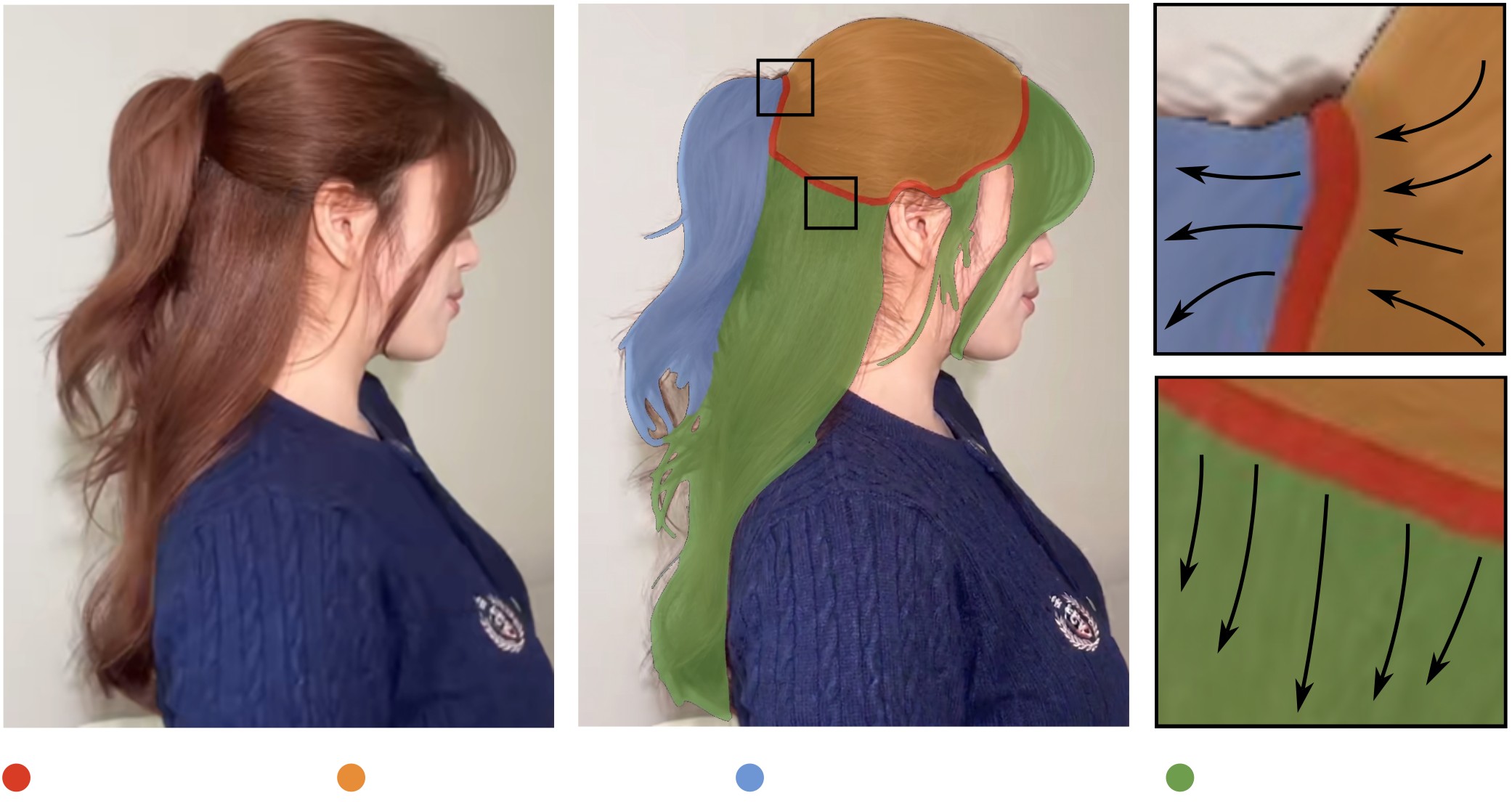}
    \put(-236,2){\footnotesize Parting line}
    \put(-182,2){\footnotesize Gathering region}
    \put(-118,2){\footnotesize Constrained region}
    \put(- 48,2){\footnotesize Loose region}
    \caption{Four semantic regions on a hairstyle with constrained structure. The close-up shows the hair flow directions within each region.}
    \label{fig:semantic}
    \Description{}
\end{figure}
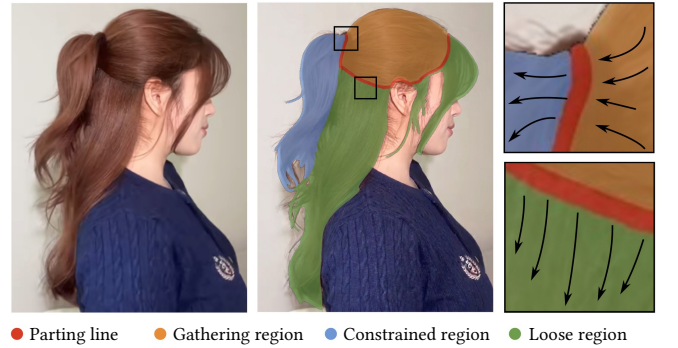

\paragraph{Hair Semantics.}
We consider four semantic concepts (\autoref{fig:semantic}) that are particularly important for hair organization:
\begin{itemize}[leftmargin=*]
    \item \emph{Parting line}: a region on the scalp where hair separates into different flow directions, often visible as a valley or split.
    \item \emph{Gathering region}: a region in which hair progressively converges toward a tied or constrained structure.
    \item \emph{Constrained region}: a region where hair is physically constrained into a compact form, such as a ponytail or a bun.
    \item \emph{Loose region}: hair that is not tied or otherwise constrained and therefore flows freely.
\end{itemize}

\paragraph{Semantic Labeling.}
From each rendered image $\mathcal{I}^{\text{render}}_i$, we ask the LMM to identify two types of semantic cues: \emph{parting lines} and \emph{constrained regions} (e.g., ponytails or buns), encoded using different colors. Prompt details are provided in the supplementary material. These labeled pixels are then projected back onto the hair surface mesh $\mathcal{M}^{\text{hair}}$.
For the projected parting labels, we first identify connected components on the scalp mesh and fit a polyline 
to each component. We then merge nearby polylines whose tangent directions differ by less than $15^\circ$ and whose endpoints are within half of the median scalp-mesh edge length. The scalp mesh provides a stable, topology-fixed spatial reference across hairstyles. We obtain constrained regions by projecting the corresponding semantic masks onto the mesh and aggregating them across views. Given the detected parting lines and constrained regions, we define each \emph{gathering region} as the surface region between a parting line and its associated constrained region. To enforce topological regularity, we apply a flood-fill to the surface and remove isolated unlabeled patches within each gathering region, yielding a topological disk. 

\paragraph{Surface Partition.}
Using the extracted semantics, we partition $\mathcal{M}^{\text{hair}}$ into four region types: \emph{loose}, \emph{gathering}, \emph{constrained}, and \emph{scalp}. 
The constrained region is directly derived from multi-view semantic masks projected onto the 3D mesh surface. Gathering regions are inferred from the spatial relationship between parting lines and constrained regions. The scalp region is inherited from the aligned bust template (\autoref{sec:image2mesh}). Any remaining surface area is classified as loose hair. For hairstyles without constrained structures, the partition reduces to only the scalp and loose regions. A hairstyle may contain multiple loose, gathering, and constrained regions.

\paragraph{Remark.}
Although the spline merging and region cleanup steps involve several thresholds, we found this semantic labeling pipeline to be robust across all examples shown in the paper. Importantly, these semantics are not used to prescribe detailed strand geometry; rather, they provide sparse but high-level structural constraints that are difficult to infer reliably from local appearance alone.

\subsection{Orientation Diffusion} \label{sec:diffusion}
\begin{figure}[t!]
 \newcommand{\figcap}[1]{\begin{minipage}{0.24\linewidth}\centering#1\end{minipage}}
    \centering
    \includegraphics[trim = 0 16.8 0 0, clip, width=\linewidth]{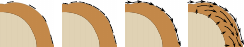} 
    \figcap{\small (a)}\hfill
    \figcap{\small (b)}\hfill
    \figcap{\small (c)}\hfill
    \figcap{\small (d)}
    \caption{Illustration of our orientation diffusion procedure. Starting from the projected directions (a), we first fill missing surface directions via surface direction diffusion (b), then resolve the sign ambiguity to obtain oriented directions on the surface (c), and finally diffuse the oriented field into the hair volume using the scalp normals (light yellow region) as boundary conditions (d).}
    \label{fig:orientation}
    \Description{}
\end{figure}

The projected tensors described above provide directional cues on visible parts of the surface, but three key challenges remain. First, directional information is only available on visible surface regions, leaving the hair interior undefined and preventing volumetric strand tracing. Second, the recovered field is an unoriented \emph{direction} field rather than an \emph{orientation} field, suffering from sign ambiguity. Third, due to occlusion and limited viewpoint coverage, some surface regions receive no directional observations. To address these issues, we introduce a three-stage diffusion pipeline as illustrated in~\autoref{fig:orientation}.

\paragraph{Data Structure.}
Previous methods~\cite{Shen2023ct2hair,zheng2023hairstep,wu2024monohair} typically represent hair direction or orientation fields on a structured voxel grid. In our setting, however, directional information is naturally attached to the vertices of the reconstructed hair surface. We therefore adopt a hybrid graph representation that treats surface vertices and volumetric samples uniformly.
Specifically, we sample the hair volume $\Omega$ using Poisson disk sampling. Let $\mathcal{S}^{\Omega} = \{ \ss_i \}$ denote the volumetric samples and $\mathcal{V}^{\text{surface}} = \{ \vvv_i \}$ the set of vertices on $\mathcal{M}^{\text{hair}}$. We define the complete sample set as $\mathcal{S}^{\text{total}} = \mathcal{S}^{\Omega} \cup \mathcal{V}^{\text{surface}}.$ We then build a $k$-nearest-neighbor graph over $\mathcal{S}^{\text{total}}$.
Not all spatially nearby samples should be connected. In particular, edges that cross semantic discontinuities would incorrectly smooth together incompatible hair flows. We therefore prune the graph using semantic barriers. First, to preserve directional discontinuities across parting lines, we remove graph edges that cross any parting spline. To ensure that this separation extends to the scalp, we prolong each parting spline to the scalp surface: for each spline point, we find its closest point on the scalp, and use the spline, the induced scalp curve, and the connecting segments to form a quad-strip barrier. 
Any graph edge intersecting is removed.
%
The resulting pruned graph is denoted by $\mathcal{G} = \{ e_j \},$ where each edge $e_j$ connects a pair of samples in $\mathcal{S}^{\text{total}}$.

\paragraph{Stage I: Surface Direction Diffusion.}
We first complete missing \emph{surface directions} while preserving the sign-invariant tensor representation. Let $\mathcal{G}^{\text{surface}} \subset \mathcal{G}$ denote the subgraph restricted to surface vertices. We optimize the tensor field by encouraging smoothness over neighboring vertices while keeping directions near the parting region fixed:
\begin{align} \label{eq:dir_diffusion}
\min_{\{\MM_i\}} \quad & \sum_{(i,j) \in \mathcal{G}^{\text{surface}}} \|\MM_i - \MM_j\|_F^2
+ w^\text{p} \sum_{\vvv_i \in \mathcal{V}^{\text{parting}}} \|\MM_i - \MM_i^0\|_F^2.
\end{align}
Here, each pair $(i,j)$ denotes an edge connecting samples $i$ and $j$ in the surface graph, $\MM_i^0$ is the initial tensor obtained from projected directions, $\mathcal{V}^{\text{parting}} \subset \mathcal{V}^{\text{surface}}$ denotes the set of surface samples located near the parting line, $\|\cdot\|_F$ denotes the Frobenius norm, and $w^\text{p} = 10^6$ is a large scalar weight used to enforce the boundary condition. Intuitively, this step fills unlabeled surface regions by diffusing reliable directional evidence across the mesh while preserving sharp directional transitions near the parting structure. In practice, we solve \autoref{eq:dir_diffusion} on the surface graph using a conjugate gradient solver. The optimization terminates when either the maximum number of iterations ($2000$) is reached or $L_2$ norm of residual falls below $1 \times 10^{-8}$.


\paragraph{Stage II: Surface Orientation Assignment.}
After Stage I, every surface vertex is associated with an unoriented direction, but the sign ambiguity remains unresolved. Prior methods often rely on global heuristics, such as aligning hair with scalp normals~\cite{paris2008hair} or assuming gravity-dominated flow~\cite{takimoto2024dr}. Such assumptions are often too weak for complex hairstyles, especially for gathered or tied styles, in which neighboring regions may exhibit opposing or even orthogonal flows.
Instead, we resolve the sign ambiguity using the semantic surface partition introduced in \autoref{sec:partition}. Our key assumption is that each semantic region exhibits a characteristic flow pattern:
\begin{itemize}[leftmargin=*]
    \item In a \emph{gathering} region, hair flows toward the interface shared with the associated constrained region.
    \item In a \emph{constrained} region, hair flows away from that interface.
    \item In a \emph{loose} region, hair flows away along the gravity direction.
\end{itemize}
These assumptions specify only the \emph{sign} of the local direction field; the local tangent direction itself is still inherited from the diffused tensors. Using these boundary cues, we first orient directions along the relevant boundary curves and then propagate the chosen signs to neighboring vertices until each non-scalp surface vertex is assigned a consistent orientation $\oo_i$.
We exclude the scalp region from this stage. At the scalp, strand roots should emerge approximately along the outward scalp normal, which serves as a more appropriate boundary condition for volumetric propagation. 

\paragraph{Stage III: Volume Orientation Diffusion.}
Finally, we propagate the oriented surface field into the hair interior. Let $\mathcal{G}^{\Omega} \subset \mathcal{G}$ denote the diffusion graph defined over both volumetric and surface samples. We solve for an orientation field $\{\oo_i\}$ by minimizing
\begin{align}
\min_{\{\oo_i\}} \quad
\sum_{(i,j) \in \mathcal{G}^{\Omega}} \|\oo_i - \oo_j\|^2
&+ w^\text{t} \sum_{\vvv_i \in \mathcal{V}^{\text{surface}} \setminus \mathcal{V}^{\text{scalp}}} \|\oo_i - \oo_i^0\|^2 \notag \\
& 
+ w^\text{n} \sum_{\vvv_i \in \mathcal{V}^{\text{scalp}}} \|\oo_i - \nn_i\|^2 \;,
\end{align}
where $\oo_i^0$ is the oriented surface direction obtained in Stage II, $\nn_i$ is the outward unit normal on the scalp, and $\mathcal{V}^{\text{scalp}} \subset \mathcal{V}^{\text{surface}}$ denotes the subset of surface vertices belonging to the scalp. Intuitively, this step produces a smooth volumetric orientation field that is anchored by the semantically disambiguated surface flow, while enforcing physically meaningful root directions in the scalp region.
%
The optimization uses the same conjugate gradient setup as in the surface case, with a maximum of $2000$ iterations and an $L_2$ residual threshold of $1 \times 10^{-8}$. Surface and scalp constraints are enforced softly using penalty weights $w^\text{t} = 10$ and $w^\text{n} = 10^{6}$.

\subsection{Strand Tracing} \label{sec:tracing}

Once the volumetric orientation field has been recovered, we generate hair strands by tracing integral curves through it. We follow the bidirectional tracing strategy of prior work~\cite{Shen2023ct2hair}.
At a query position $\xx$, we estimate the local orientation by blending nearby samples $\ss_i$ within a kernel of radius $r$, where $r$ is set to the average nearest-neighbor distance in $\mathcal{S}^{\Omega}$.
\begin{align} 
\oo(\xx) =
\text{\fontfamily{lmss}\selectfont normalize}
\Bigl(
{\sum_i w_i \oo_i} / {\sum_i w_i}
\Bigr),
\end{align}
where $w_i$ is an inverse-distance weight determined by the distance between $\xx$ and the sample location $\ss_i$.

\paragraph{Forward Tracing.}
We first seed strands on the scalp and advect them along the orientation field, following prior work~\cite{Chai2013,shen2020deepsketchhair,zheng2023hairstep,zhou2024groomcap}. Tracing terminates when a strand reaches the boundary of the hair volume. This stage produces the primary set of strands that originate from plausible root locations.

\paragraph{Gap Filling.}
Forward tracing alone may leave sparsely covered regions in the interior of the hairstyle. To improve spatial coverage, we examine each volumetric sample $\ss \in \mathcal{S}^{\Omega}$ after forward tracing. If no strand passes through the sphere of radius $r$ centered at that sample, we initialize an additional strand at that location and trace it in both directions along the orientation field. Tracing stops when one direction reaches the scalp, and the other reaches the hair boundary. This gap-filling stage increases strand coverage while remaining consistent with the recovered volumetric flow.

\section{Results}

All experiments are conducted on a single GPU with 5,120 cores and 640 Tensor cores, 32GB memory, and the entire pipeline is implemented in Python. We use Tripo 3D~\cite{Tripo3} as the large reconstruction model (LRM) for image-to-mesh reconstruction and Nano Banana 2~\cite{nanobanana2026} as the large multimodal model (LMM) for grayscale guidance generation, parting line detection, and constrained region segmentation. As our method requires no task-specific training, these backbone models can be readily replaced by other alternative foundation models. $|\mathcal{S}^\text{total}|$ is around 10k samples over all examples. Please refer to the appendix for our prompt.

\begin{figure*}[ht!]
\centering
\includegraphics[width=0.75\linewidth]{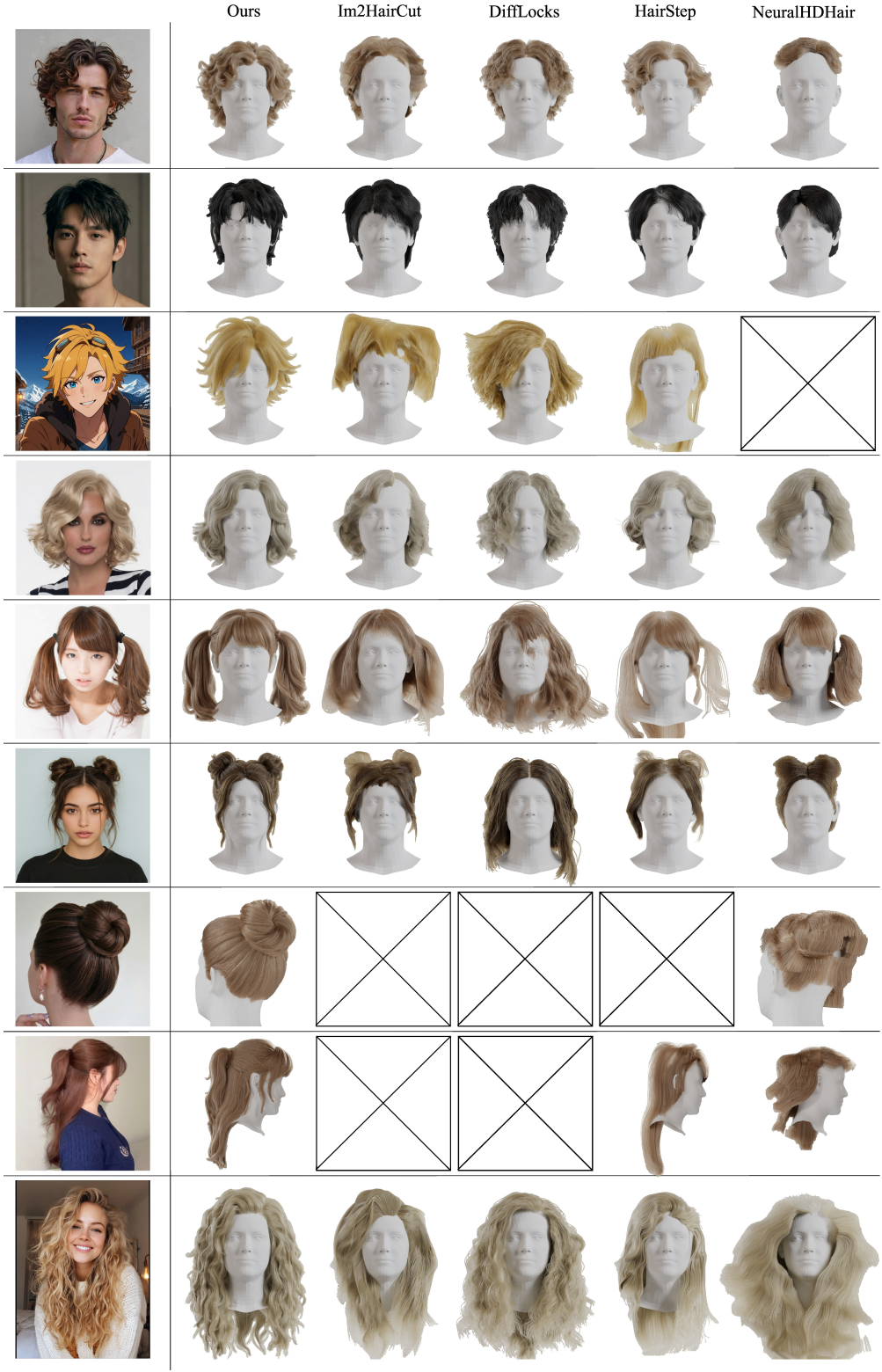}
\caption{Visual comparison with Im2HairCut~\cite{sklyarova2025im2haircut}, DiffLocks~\cite{difflocks2025}, HairStep~\cite{zheng2023hairstep}, and NeuralHDHair~\cite{wu2022neuralhdhair} across a variety of hairstyles. Crossed-out entries indicate that the corresponding method fails to produce a result for that hairstyle.}
\label{fig:validation_result}
\Description{}
\end{figure*}

\begin{figure*}[ht!]
    \centering

    \noindent
    \vspace{1mm}

    \includegraphics[
        width=0.8\linewidth
    ]{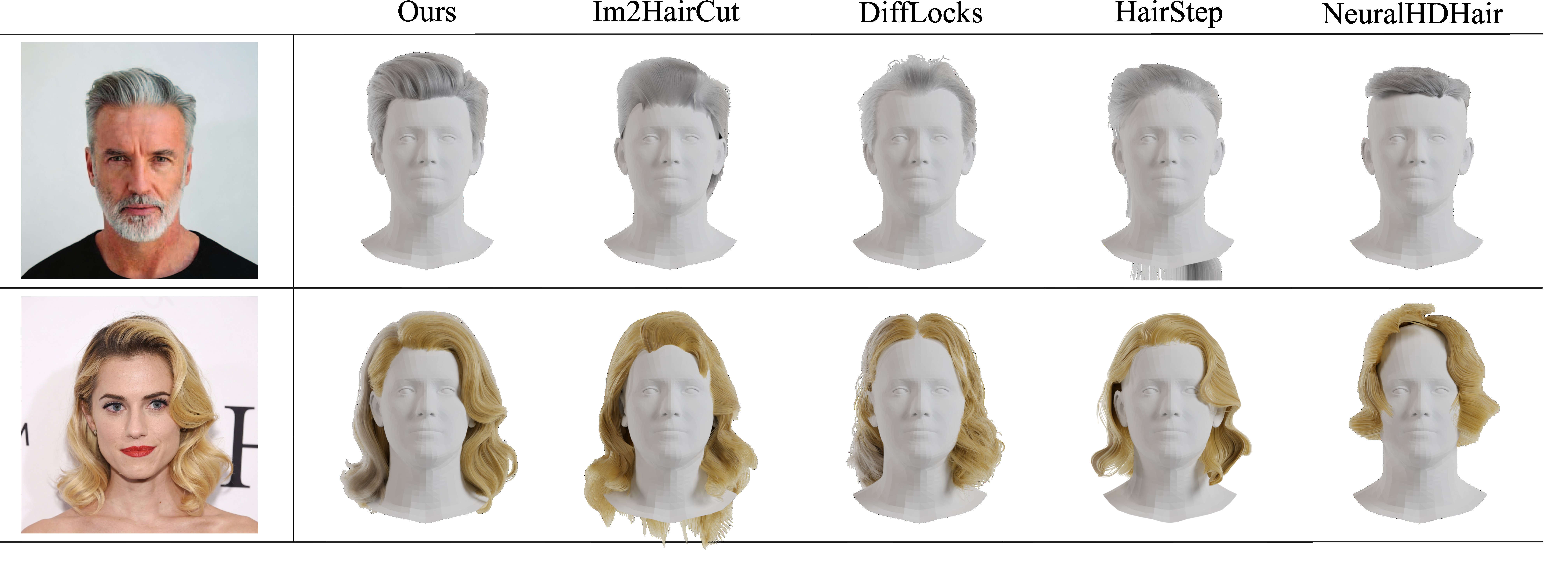}

    \caption{
        Additional visual comparisons with state-of-the-art methods.
        Our method consistently produces more coherent strand geometry
        and faithful hair shape across all examples.
    }
    \label{fig:validation_sup}
    \Description{}
\end{figure*}

\paragraph{Comparison with SOTAs}
We evaluate our method through qualitative comparisons on in-the-wild images and ablation studies that validate the key components of our pipeline. We compare against four state-of-the-art single-image hair reconstruction methods: NeuralHDHair~\cite{wu2022neuralhdhair}, HairStep~\cite{zheng2023hairstep}, DiffLocks~\cite{difflocks2025}, and Im2Haircut~\cite{sklyarova2025im2haircut}. As shown in~\autoref{fig:validation_result} and ~\autoref{fig:validation_sup}, our method robustly reconstructs 3D strand-based hair models from a single image, even when only a side view or a back view is available. Our approach also generalizes well to a wide range of hairstyles, including straight, wavy, and curly hair; short and long styles; and challenging cases such as buns and ponytails. Benefiting from LRMs, our method captures the overall hair shape and silhouette more accurately. LMMs further improve the prediction of surface hair directions and parting lines. In addition, our three-stage orientation diffusion scheme produces smoother and more coherent strand propagation throughout the hair volume. In contrast, prior learning-based methods, limited by the diversity of their training data, often fail to faithfully recover hair geometry and surface orientation, especially for unseen hairstyles such as ponytails and buns. 

\begin{figure*}[ht!]
\centering
\newcommand{\figcap}[1]{\begin{minipage}{0.16\linewidth}\centering#1\end{minipage}}
\figcap{\hspace{4mm}Input}\hfill
\figcap{\small\hspace{5mm} HairLRM}\hfill
\figcap{\small\hspace{5mm} Ours}\hfill
\figcap{\small\hspace{4mm} Input}\hfill
\figcap{\small\hspace{2mm} HairGPT}\hfill
\figcap{\small Ours}\hfill
\includegraphics[width=\linewidth]{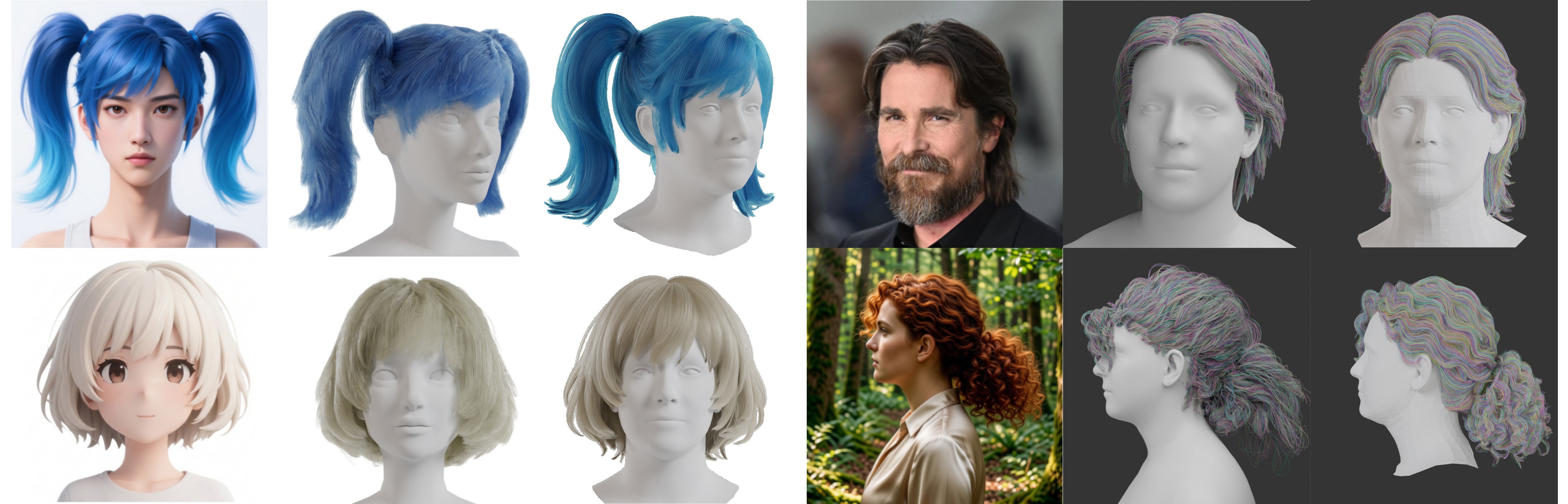}
\caption{Compared with these recent approaches, our method better preserves overall hairstyle shape and local geometric details, benefiting from the high-quality mesh provided by the LRM and our dedicated three-stage reconstruction pipeline.
}
\label{fig:comp_sota}
\Description{}
\end{figure*}

To further quantitatively compare our results with prior work, we follow Im2Haircut~\cite{sklyarova2025im2haircut} by manually aligning all 3D reconstructions to the ground truth before evaluation. We then render each reconstruction from the input view and compare it against the corresponding input image. We use Intersection over Union (IoU) to measure hair silhouette overlap in the input view, and Ori-Err to measure the mean angular error between the 2D orientations extracted by HairStep~\cite{zheng2023hairstep} from the rendered reconstruction and from the ground truth. \autoref{tab:metrics_2d} reports the quantitative results. Our method achieves the highest IoU (0.75) and the lowest orientation error ($32.6^\circ$), substantially outperforming all competing approaches on both metrics. These results indicate that our method achieves the highest quantitative accuracy and produces the most visually faithful single-view reconstruction results.
\begin{table}[ht]
\centering
\caption{Quantitative comparison with prior methods. }
\label{tab:metrics_2d}
\renewcommand{\arraystretch}{1.15}
\scalebox{0.9}{
\begin{tabular}{l|cc}
    \toprule
    {Method} & {IoU} $\uparrow$ & {Ori-Err} $\downarrow$ \\
    \midrule
    NeuralHDHair~\cite{wu2022neuralhdhair}   & 0.52 & $68.4^{\circ}$ \\
    HairStep~\cite{zheng2023hairstep}        & 0.64 & $55.5^{\circ}$ \\
    DiffLocks~\cite{difflocks2025}       & 0.61 & $55.6^{\circ}$ \\
    Im2Haircut~\cite{sklyarova2025im2haircut}& 0.64 & $45.1^{\circ}$ \\
    \textbf{Ours}                            & \textbf{0.75} & $\mathbf{32.6^{\circ}}$ \\
    \bottomrule
  \end{tabular}
  }
\end{table}

We further compare with the latest learning-based hair reconstruction methods, HairLRM~\cite{shen2026HairLRM} and HairGPT~\cite{luo2026}. As neither is open-sourced, we use the same image from their paper as input and directly compare it with their rendered results. As shown in~\autoref{fig:comp_sota}, our method better preserves overall hairstyle shape and local geometric details, benefiting from the high-quality mesh provided by the LRM and our dedicated three-stage reconstruction pipeline.

\paragraph{Performance}
Our pipeline takes around 20 minutes per model. It begins with LRM-based mesh reconstruction, which produces a mesh with approximately 1M faces in about 3 minutes per portrait. The main runtime bottleneck is LMM inference, which takes approximately 3 minutes per portrait due to multiple queries across six rendered views for direction projection and semantic partitioning. The remaining computation is relatively lightweight. Semantic processing, including 2D-to-3D direction projection, partition wall construction, and region labeling, takes approximately 3 minutes. Our core three-stage orientation diffusion pipeline also takes approximately 3 minutes, and the subsequent strand tracing requires about 8 minutes for 50{,}000 strands. For reference, NeuralHDHair, DiffLocks, and Hairstep process a single image in about 2, 4, and tens of minutes, respectively, for 10{,}000 strands. Im2Haircut requires hours due to differentiable optimization. Note that we have not tuned the output mesh or image resolutions used by the LRMs and LMMs, and the tracing stage has not been optimized either, leaving substantial room for performance improvement.

\begin{figure}[ht!]
\centering
\newcommand{\figcap}[1]{\begin{minipage}{0.495\linewidth}\centering#1\end{minipage}}
\includegraphics[trim = 300 0 0 0, clip,width=\linewidth]{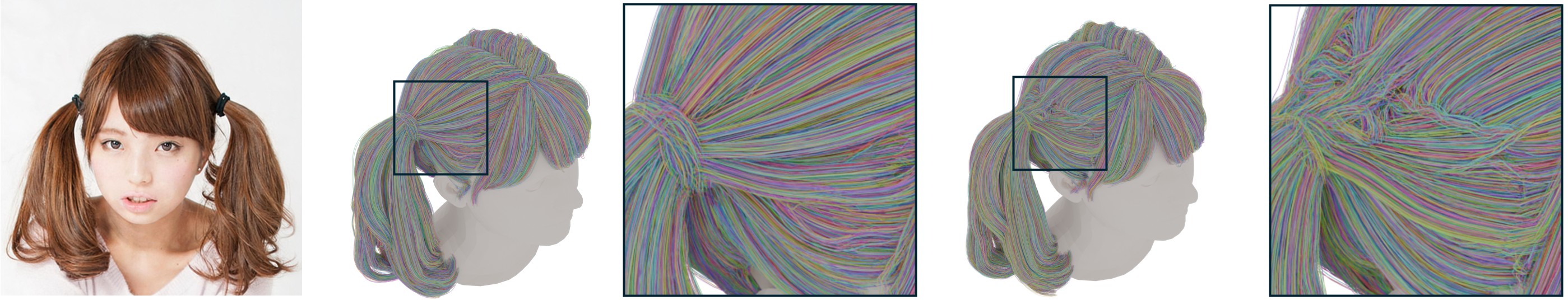}
\figcap{\small {Ours}} 
\figcap{\small HairStep + Dr.~Hair}\hfill
\caption{Our method produces smooth and coherent hair strands, whereas HairStep + Dr.~Hair yields locally disordered hair directions due to the limited manually annotated hair-orientation training set.}
\vspace{-0.3cm}
\label{fig:ablation_hairstep}
\Description{}
\end{figure}

\paragraph{Compared with orientation prediction followed by volume diffusion.}
To validate the necessity of our three-stage orientation diffusion pipeline, we first compare against an alternative pipeline that predicts hair orientation using HairStep~\cite{zheng2023hairstep} and then diffuses the orientation throughout the hair volume following Dr.~Hair~\cite{takimoto2024dr}. As shown in \autoref{fig:ablation_hairstep}, HairStep generalizes poorly to unseen hairstyles due to its limited manually annotated training set on hair orientation. In particular, it fails to produce reliable orientations in challenging regions such as gathering and parting areas. As a result, the subsequent volume diffusion yields locally disordered hair directions, even when our parting boundary conditions are applied.

\begin{figure}[ht!]
\centering
\newcommand{\figcap}[1]{\begin{minipage}{0.495\linewidth}\centering#1\end{minipage}}
\includegraphics[trim = 0 0 0 0, clip,width=\linewidth]{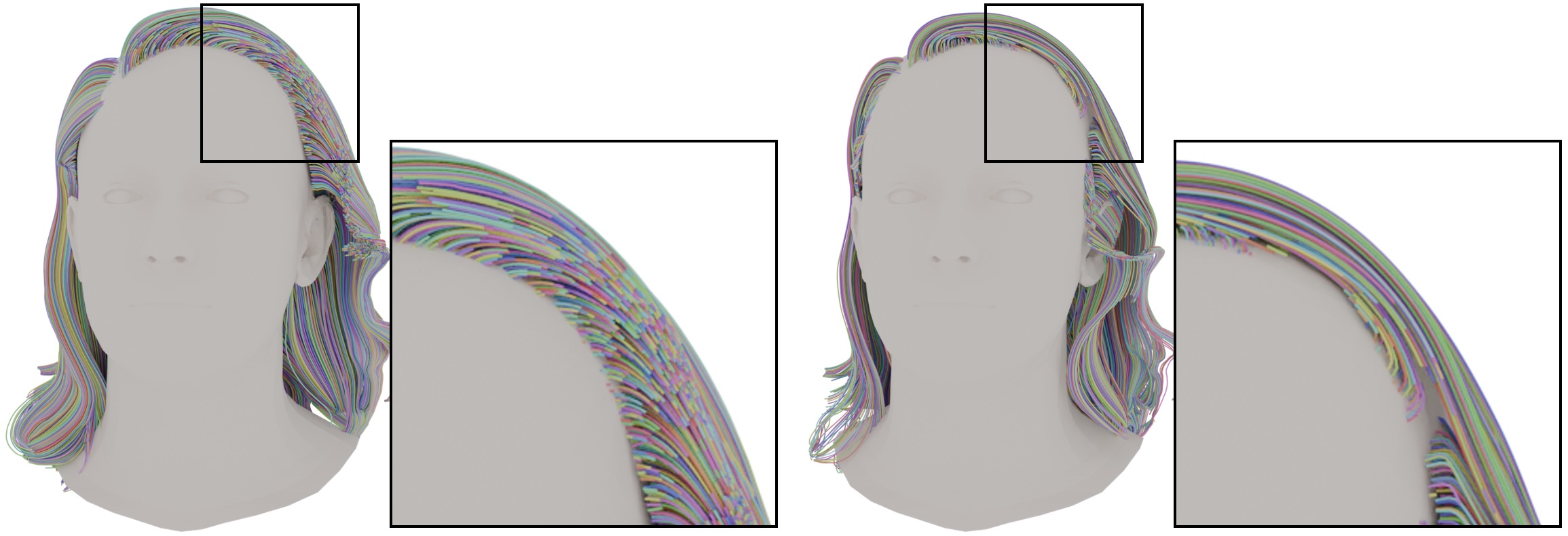}
\figcap{\small {Ours}} 
\figcap{\small Volumetric Direction Diffusion}\hfill
\caption{Diffusing unoriented directions in the volume leads to strands that loop inward toward the scalp due to directional ambiguity, whereas our method produces smooth and coherent hair strands.}
\vspace{-0.3cm}
\label{fig:ablation_oridiffusion}
\Description{}
\end{figure}

\paragraph{Compared with unoriented direction diffusion.}
We further compare against a baseline that extracts unoriented hair directions on the hair surface using Gabor filter analysis~\cite{jakob2009capturing,Paris2004} and then directly diffuses these directions throughout the hair volume. As shown in \autoref{fig:ablation_oridiffusion}, diffusing unoriented directions in the volume leads to strands that loop inward toward the scalp due to directional ambiguity, whereas our method produces smooth and coherent hair strands.

\begin{figure*}[ht!]
\centering
\includegraphics[width=\linewidth]{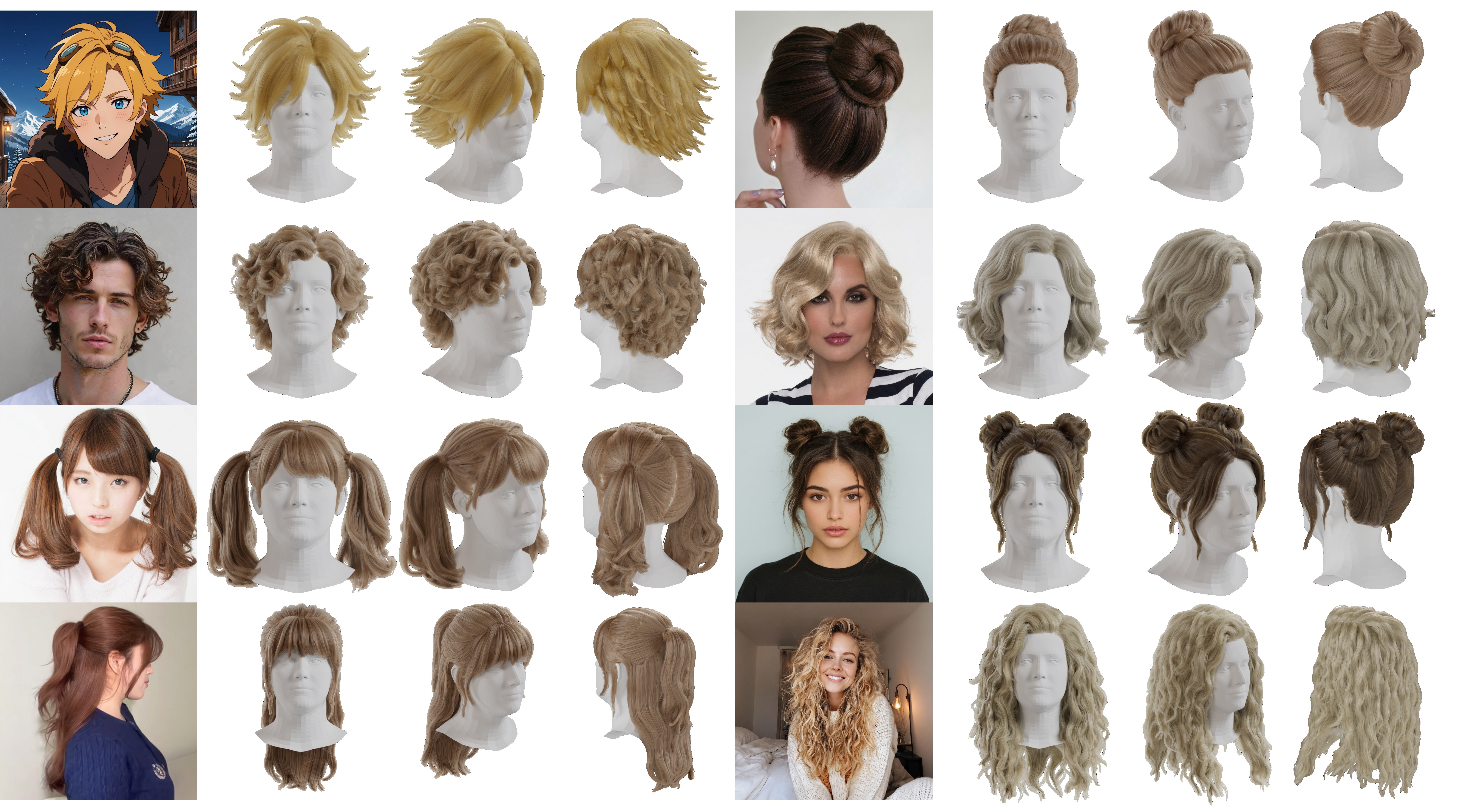}
\caption{Different views of our results, demonstrating the globally consistent 3D geometry of our reconstructed strands. 
}
\label{fig:multiview}
\Description{}
\end{figure*}

To further demonstrate that our method recovers globally consistent 3D geometry rather than merely fitting the input viewpoint, \autoref{fig:multiview} presents our reconstructions rendered from multiple surrounding views. The recovered strand geometry remains coherent and visually plausible across all viewpoints, confirming the effectiveness of our volumetric orientation diffusion in producing a complete and consistent 3D strand field.

\begin{figure}[ht!]
\centering
\includegraphics[width=\linewidth]{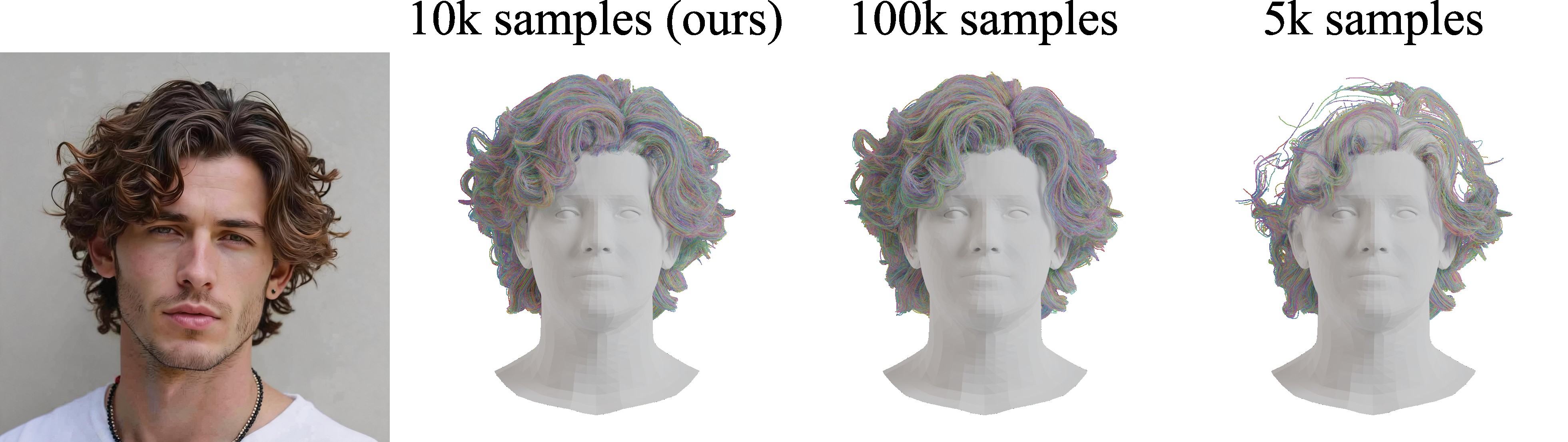}
\caption{Ablation on sampling density. 5k samples cause strand tracing to fail, while 100k samples show no significant improvement over our default 10k. }
\vspace{-0.3cm}
\label{fig:sampling_density}
\Description{}
\end{figure}

\paragraph{Ablation on Sampling Density}
We ablate the effect of volumetric sampling density on reconstruction quality in~\autoref{fig:sampling_density}.  Reducing the sample count to 5k causes strand tracing to fail due to insufficient interior samples for reliable orientation interpolation. 100k samples yield results with no significant visual improvement compared with ours, while requiring more computation time for orientation diffusion. Our default setting achieves a good balance between reconstruction quality and efficiency.

\begin{figure*}[ht!]
\centering
\newcommand{\figcap}[1]{\begin{minipage}{0.495\linewidth}\centering#1\end{minipage}}
\includegraphics[trim=380 150 400 130,clip,width=0.332\linewidth]{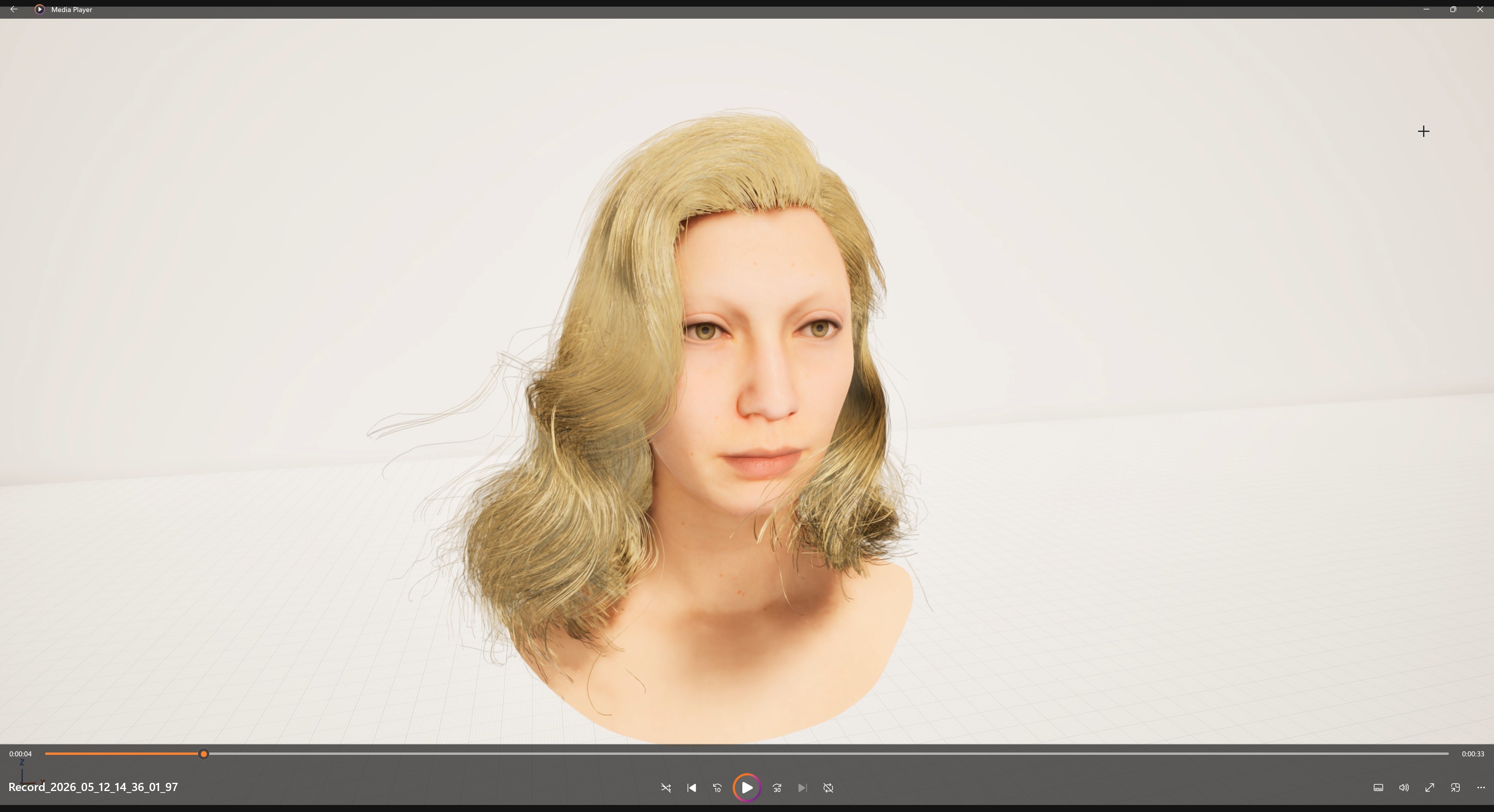}\hfill
\includegraphics[trim=380 150 400 130,clip,width=0.332\linewidth]{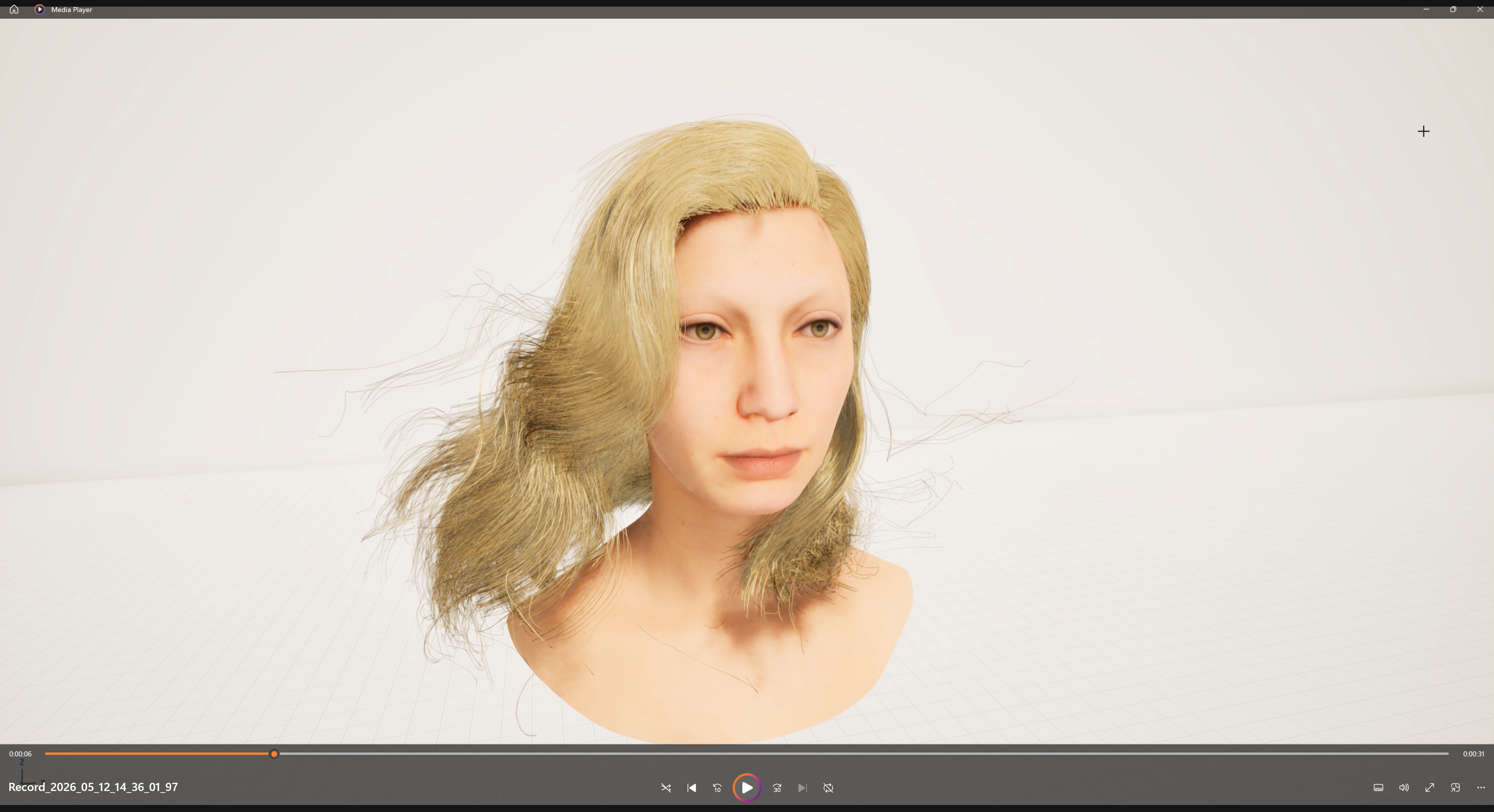}\hfill
\includegraphics[trim=380 150 400 130,clip,width=0.332\linewidth]{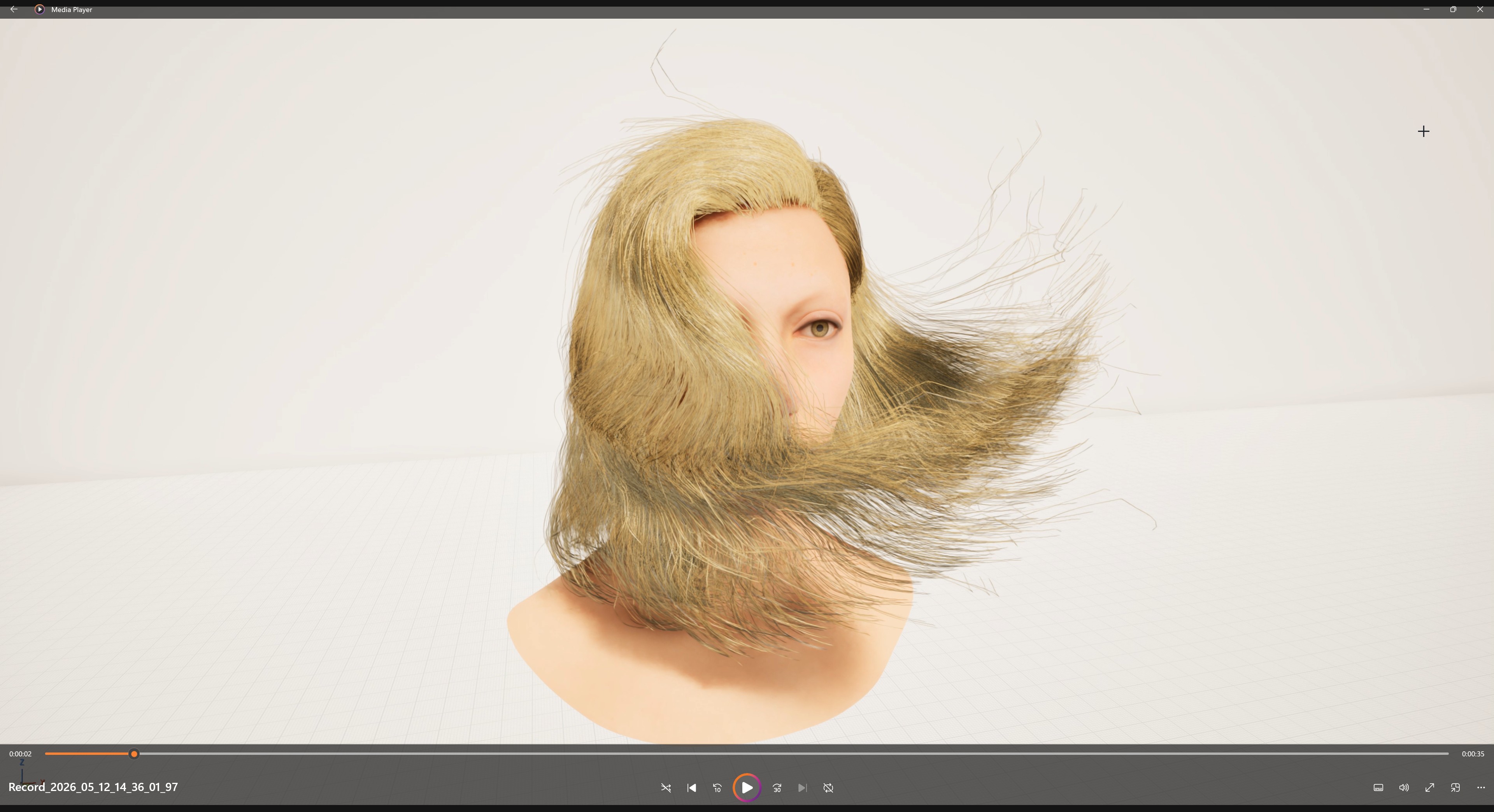}
\caption{Importing our reconstructed hair strands into Unreal Engine~5 to demonstrate strand-based simulation and rendering capabilities.}
\label{fig:UE}
\Description{}
\end{figure*}

\begin{figure*}[ht!]
\centering
\includegraphics[width=\linewidth]{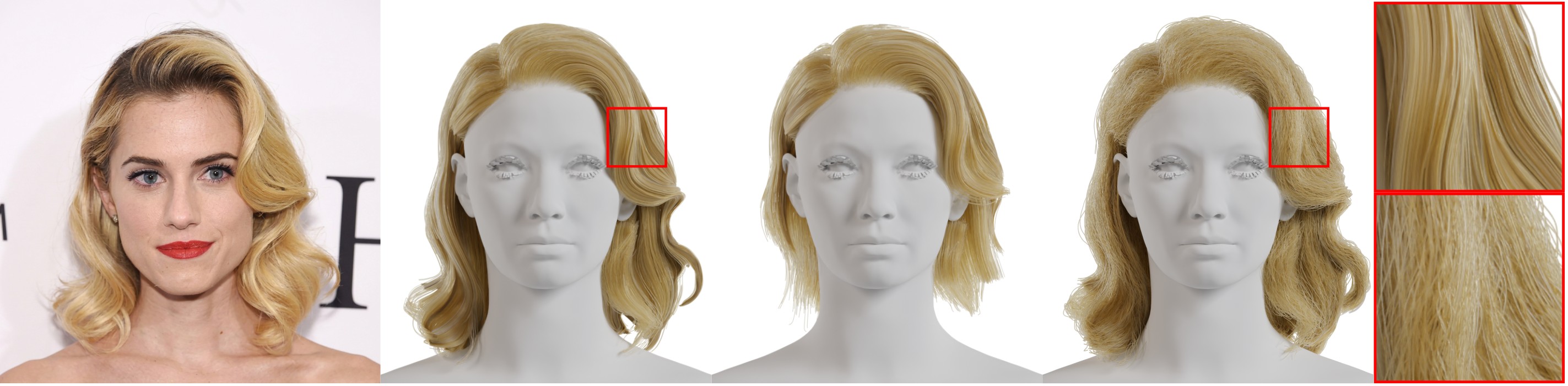}
\put(-467,-7){\small Input image}
\put(-388,-7){\small Our reconstructed strand-based model}
\put(-254,-7){\small Adding scale modifier}
\put(-149,-7){\small Adding fuzzy modifier}
\caption{Example of applying scale and fuzzy groom modifier to our reconstructed strand-based hairstyle}
\label{fig:groom}
\Description{}
\end{figure*}

\paragraph{Applications}
We further demonstrate two downstream applications of our strand-based hair reconstruction. First, \autoref{fig:UE} demonstrates that our reconstructed hair strands can be directly imported into the industrial game engine Unreal Engine~\cite{unrealengine} to enable strand-based simulation and rendering. Second, our representation remains compatible with standard grooming workflows, allowing artists to apply grooming operators for secondary editing, enrich fine-scale details, and create style variations such as using scale and fuzz modifiers shown in~\autoref{fig:groom}.

\section{Conclusion}

\begin{figure*}[h]
\centering
\includegraphics[width=\linewidth]{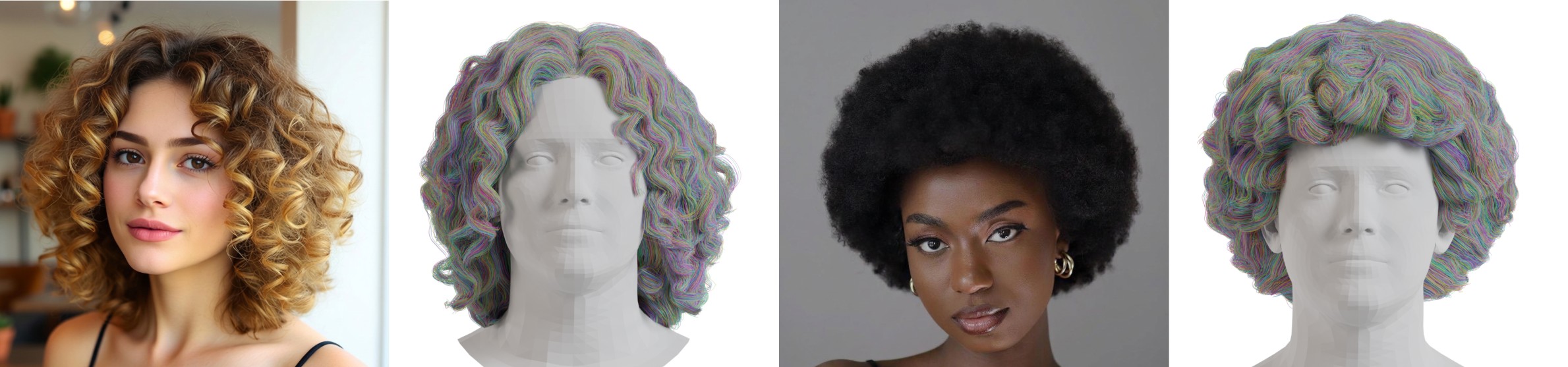}
\caption{Failure cases of our method on highly intricate hairstyles. Complex strand interactions and ambiguous local structures may lead to inaccurate strand flow reconstruction, missing fine-scale details, and locally inconsistent geometry.}
\label{fig:failure}
\Description{}
\end{figure*}

We have presented a novel pipeline for strand-based hairstyle reconstruction that combines large multimodal models, large reconstruction models, and conventional graphics techniques. From a single input image, our method reconstructs a coarse hair surface, infers surface directions and structural regions, and then recovers a consistent volumetric orientation field through a diffusion-correction-diffusion process. Hair strands are finally generated by tracing through the recovered field. Our method supports a wide range of hairstyles, including straight, curly, short, and long styles, as well as more complex styles such as buns and ponytails. Compared with previous approaches, it requires neither training nor data collection and produces high-quality strand-based results efficiently within a few minutes.
Overall, our work suggests that combining foundation models with classical geometry processing provides a practical and effective direction for strand-based hair modeling and offers a new path toward scalable reconstruction of complex hairstyles from minimal input.

\paragraph{Limitations and Future works}
Despite the promising results, our method still exhibits some limitations. In particular, highly intricate hairstyles, especially those with curly or kinky hair (\autoref{fig:failure}), remain challenging because their complex interwoven topology is difficult to infer reliably from a single image. In addition, fine-scale strand details that are only weakly represented in the reconstructed surface are also difficult to recover faithfully. These limitations suggest several promising directions for future work. Extending the framework to better support braids and other complex topological structures is an important next step. Additionally, tighter integration between geometric reconstruction and semantic hairstyle understanding may improve robustness for complex strand topologies. Interactive editing tools also represent a promising direction for enabling user-guided refinement and greater artistic control over the generated hairstyles.

\newpage

\bibliographystyle{ACM-Reference-Format}
\bibliography{ref}

\begin{figure*}[t!]
\centering
\includegraphics[width=\linewidth]{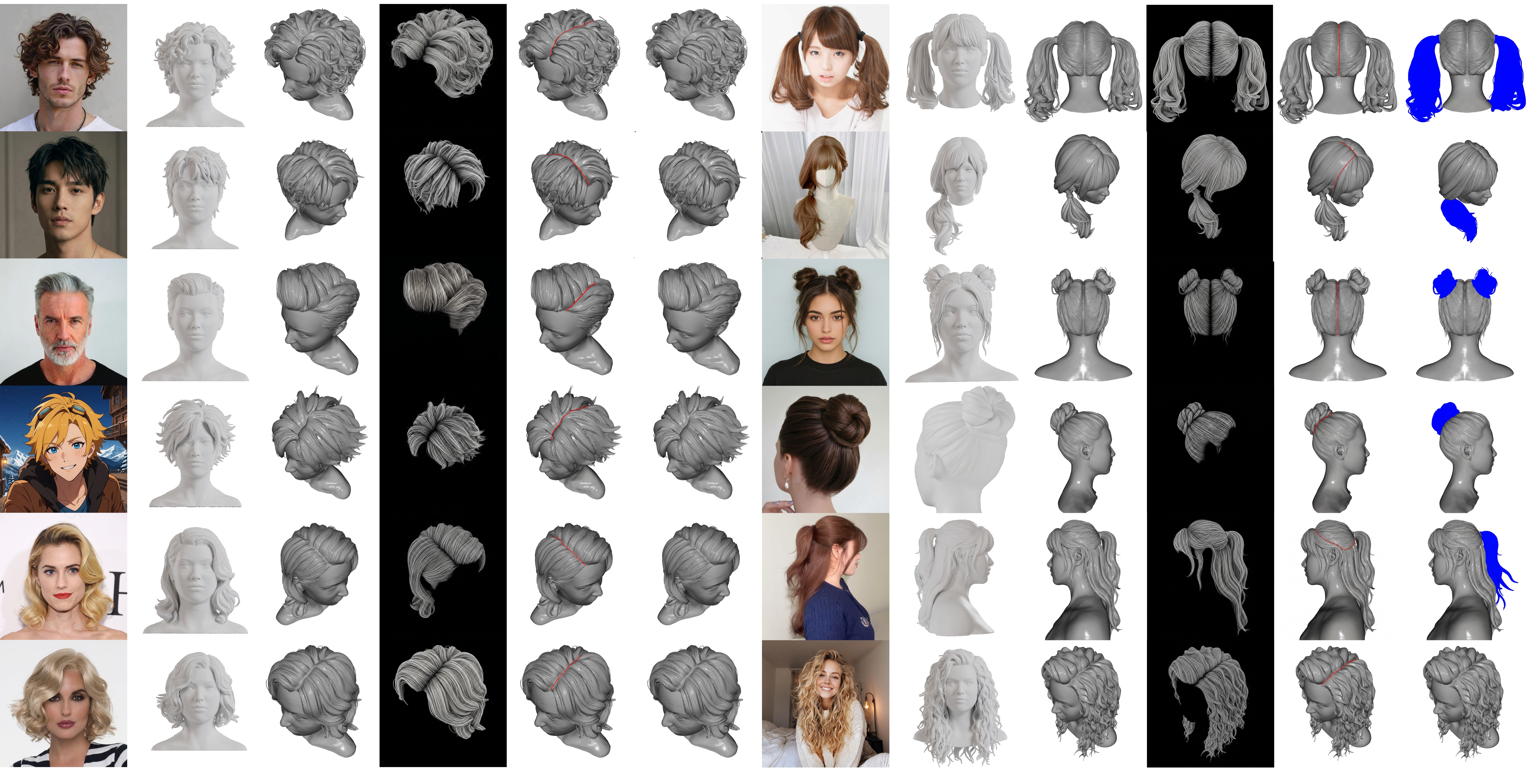}
\caption{Detailed per-step results of our LMM queries. Each group shows, from left to right: the input image, the reconstructed mesh, the rendered image, and the corresponding LMM outputs including grayscale guidance, parting line annotation (red), and constrained region mask (blue).}
\label{fig:prompt_result}
\Description{}
\end{figure*}

\newpage

\appendix
\section{Prompts}
We leverage an LMM to extract semantic information from multi-view rendered images of the reconstructed hair mesh. Specifically, the LMM is queried for three complementary tasks: generating grayscale guidance images that emphasize local strand flow for direction estimation, detecting parting lines that indicate where hair separates into different flow directions, and identifying constrained regions such as ponytails and buns. The full prompts used for each task are provided below, and detailed per-step results are shown in \autoref{fig:prompt_result}.
{\subsection{Render image 2 grayscale texture prompt}
\ttfamily Process this hair image into a clean grayscale texture map optimized for gradient-based orientation analysis. Requirements: Remove all specular highlights and shading variations — the output should show ONLY the fine strand level texture, with no broad illumination gradients, no glossy reflections, no color cast. Convert to grayscale where brightness represents purely the local hair strand texture: bright = strand ridge, dark = gap between strands. Uniform contrast everywhere — strands in shadow regions should be equally visible as strands in highlight regions. No area should be washed out or too dark. Preserve the exact spatial structure — do NOT hallucinate, add, remove, or shift any strand. The pixel-level position of every strand must remain identical to the input. The body/scalp region and white background should become pure black (zero value). Output at the same resolution as input (1024×1024 or higher). Think of this as creating a "height map" of the hair surface texture, where each individual strand creates a thin bright ridge. The result should look like a clean SEM (scanning electron microscope) image of hair strands — pure texture, no lighting artifacts.
}

{\subsection{Parting line labeling prompt}
\ttfamily This is a grayscale render of a 3D hair mesh. Identify the single most prominent parting line in this hairstyle. A parting line is a narrow ridge or crease on the hair surface where hair diverges to flow in two clearly different directions — it is where the scalp would be visible on a real head. It typically runs as a continuous line from the front/top of the head toward the back or side. Draw exactly ONE very thin, continuous, bright red line (\#FF0000, 1 pixel wide, solid, anti-aliasing off) along this parting line. Rules: Draw exactly 1 line. No more. The line must have sculpted hair on BOTH sides. If one side is background, skin, or empty space, it is a silhouette edge — do NOT draw it. Do NOT mark the outer contour/silhouette of the hair, the hairline, or any boundary between hair and non-hair areas. Do NOT mark boundaries between overlapping hair layers or bundles that flow in similar directions. Trace the full visible length of the parting line smoothly without truncating early. If no clear parting line exists in this view, draw nothing. Do not modify any part of the original image. Only add the red overlay line on top.
}

{\subsection{Ponytail masking prompt}
\ttfamily Identify and mask gathered hair structures only (braids, buns, ponytails, tied sections). Replace these bound parts entirely with a solid blue mask (RGB: 0, 0, 255) to clearly indicate their original occupied regions. Keep all other hair regions exactly as they are, including any tightened or pulled areas near the scalp. Only the gathered/bound extensions beyond the tie points should be covered by the blue mask. Blue mask requirements: Solid, opaque fill with uniform blue color (hex \#0000FF). The mask must precisely conform to the silhouette/outline of the removed gathered hair. No feathering, transparency, or blending at mask edges — maintain hard, clean boundaries. No inpainting, background reconstruction, or content generation in masked areas. 
}

\section{Per-step results}

\autoref{fig:prompt_result} demonstrates per-step results of our method and \autoref{fig:multiview} shows more views of our results.

\end{document}